\documentclass[12pt]{article}
\usepackage{amsmath}
\usepackage[dvipsnames]{xcolor}
\usepackage{graphicx}
\usepackage{enumerate}
\usepackage{natbib}
\usepackage{url} 
\usepackage{binomexp}
\usepackage{mathtools}
\usepackage{amsfonts}
\usepackage{mathrsfs}
\RequirePackage{subcaption}
\usepackage{bbm}
\usepackage{dsfont}
\usepackage{longtable}
\usepackage{yhmath}
\usepackage{bm}
\usepackage{amssymb}
\usepackage[nodisplayskipstretch]{setspace}
\usepackage{tikz}
\usepackage{float}
\usepackage{color}
\usepackage{stmaryrd}
\usepackage{booktabs}
\usepackage{paralist}
\usepackage{multirow}
\usepackage{threeparttable}
\usepackage{soul}
\usepackage{algorithm,algpseudocode}
\makeatletter
\newcommand{\algmargin}{\the\ALG@thistlm}
\makeatother
\newlength{\whilewidth}
\algdef{SE}[parWHILE]{parWhile}{EndparWhile}[1]
{\parbox[t]{\dimexpr\linewidth-\algmargin}{%
		\hangindent\whilewidth\strut\algorithmicwhile\ #1\ \algorithmicdo\strut}}{\algorithmicend\ \algorithmicwhile}%
\algnewcommand{\parState}[1]{\State%
	\parbox[t]{\dimexpr\linewidth-\algmargin}{\strut #1\strut}}

\newcommand{\bfx}{\bm{x}}

\newcommand{\rmd}{\mathrm{d}}
\newcommand{\expit}{\mathrm{expit}}

\newcommand{\tr}{\mathrm{tr}}
\newcommand{\dml}{\mathrm{dml}}

\newcommand{\bfX}{\bm{X}}

\newcommand{\bbE}{\mathbb{E}}

\newcommand{\bbP}{\mathbb{P}}

\newcommand{\bbU}{\mathbb{U}}

\newcommand{\bbone}{\mathbbm{1}}

\newcommand{\calB}{\mathcal{B}}

\newcommand{\calI}{\mathcal{I}}
\newcommand{\calL}{\mathcal{L}}
\newcommand{\calM}{\mathcal{M}}
\newcommand{\calN}{\mathcal{N}}
\newcommand{\calO}{\mathcal{O}}
\newcommand{\calP}{\mathcal{P}}

\newcommand{\calX}{\mathcal{X}}
\newcommand{\calY}{\mathcal{Y}}

\newcommand\independent{\protect\mathpalette{\protect\independenT}{\perp}}
\def\independenT#1#2{\mathrel{\rlap{$#1#2$}\mkern2mu{#1#2}}}

\usepackage[colorlinks=true,linkcolor=blue,citecolor=black]{hyperref}

\newtheorem{theorem}{Theorem}
\newtheorem{assumption}{Assumption}

\newtheorem{remark}{Remark}
\newtheorem{example}{Example}

\newcommand{\blind}{1}

\allowdisplaybreaks

\usetikzlibrary{shapes,decorations,arrows,calc,arrows.meta,fit,positioning}
\tikzset{
	-Latex,auto,node distance =1 cm and 1 cm,semithick,
	state/.style ={ellipse, draw, minimum width = 0.7 cm},
	point/.style = {circle, draw, inner sep=0.04cm,fill,node contents={}},
	bidirected/.style={Latex-Latex,dashed},
	el/.style = {inner sep=2pt, align=left, sloped}
}

\begin{document}

	\def\spacingset#1{\renewcommand{\baselinestretch}%
		{#1}\small\normalsize} \spacingset{1}

	
	\if1\blind
	{
		\title{\Large \bf Principal stratification with U-statistics under principal ignorability}
		\author{Xinyuan Chen$^{1,\ast}$ and Fan Li$^{2,3,\dagger}$\vspace{0.2cm}\\
			$^1$Department of Mathematics and Statistics,\\ Mississippi State University, MS, USA\\
			$^2$Department of Biostatistics, Yale School of Public Health, CT, USA\\
			$^3$Center for Methods in Implementation and Prevention Science,\\ Yale School of Public Health, CT, USA\\
			${}^\ast$xchen@math.msstate.edu\\
			${}^\dagger$fan.f.li@yale.edu}  
		\maketitle
	} \fi
	
	\if0\blind
	{
		\bigskip
		\bigskip
		\bigskip
		\begin{center}
			{\Large \bf Principal stratification with U-statistics under principal ignorability}
		\end{center}
		\medskip
	} \fi
	
	\bigskip
	\begin{abstract}
		Principal stratification is a popular framework for causal inference in the presence of an intermediate outcome. While the principal average treatment effects are the standard target of inference, they may be insufficient when interest lies in the relative ordering of potential outcomes within a principal stratum. We introduce the principal generalized causal effect estimands to accommodate nonlinear contrast functions, providing robust, probability-scale summaries suitable for ordinal outcomes and win-loss comparisons with composite endpoints. Under principal ignorability, we expand the theoretical results in \citet{Jiang2022} to a broader class of causal estimands in the presence of a binary intermediate variable. We develop nonparametric identification results and derive efficient influence functions for the generalized causal estimands in principal stratification analyses. These efficient influence functions motivate multiply robust estimators and lay the ground for obtaining efficient debiased machine learning estimators via cross-fitting based on U-statistics. The proposed methods are illustrated through simulations and the analysis of a data example. 
	\end{abstract}
	
	\noindent%

	\spacingset{1.75} 
	
	\section{Introduction}
	
	The average causal effect---defined as the expected contrast between two potential outcomes under alternative treatment conditions---is often the standard estimand of interest in causal inference. Specifically, let $\{Y(1),Y(0)\}$ denote a pair of potential outcomes for a randomly selected individual from a super-population, and then the average causal effect estimand is defined as $\bbE\{Y(1)-Y(0)\}$. Conceptually, the average causal effect estimand can be extended to a broader class, enabling arbitrary comparisons. That is, given a contrast function $h(u,v)$ defined on the product space, $\calY\times\calY$ ($\calY$: the potential outcome space), a generalized estimand can be defined as $\bbE[h\{Y(1),Y(0)\}]$. Examples include the average causal effect when $h(u,v)=u-v$, and the proportion of the
	population benefiting from treatment, $\bbP\{Y(1)> Y(0)\}$, when $h(u,v)=\bbone(u> v)$. Because $Y(1)$ and $Y(0)$ are never observed simultaneously, $\bbE[h\{Y(1),Y(0)\}]$ is generally inestimable for nonlinear $h$ (see \citet{lu2020sharp} for an example for bounding such an estimand with ordinal outcomes). However, a closely related quantity, the \emph{generalized causal effect}, can be formulated:
	\begin{align} \label{eq:sp-estimand}
		\tau_h = \int\int h(u,v) \rmd F_1(u)\rmd F_0(v),
	\end{align}
	where $F_z(\cdot)$ is the marginal distribution of the potential outcome $Y(z)\in\calY$ for $z=0,1$. Let $\{Y_1(1),Y_1(0)\}$ and $\{Y_2(1),Y_2(0)\}$ denote two independent replicates of $\{Y(1),Y(0)\}$, and $\tau_h$ can be re-expressed in a familiar form, $\bbE[h\{Y_1(1),Y_2(0)\}]$. 
	
	Two notable examples of the generalized causal effect are the probabilistic index (the target parameter for the Mann-Whitney statistic), $\bbP\{Y_1(1)\geq Y_2(0)\}$, when $h(u,v)=\bbone(u\geq v)$ \citep{vermeulen2015increasing}, and the win ratio, $\bbP\{Y_1(1)>Y_2(0)\}/\bbP\{Y_1(1)<Y_2(0)\}$, when $h(u,v)=\{\bbone(u>v),\bbone(u<v)\}$ \citep{Pocock2012}. These estimable quantities are alternative measurements for the population effect and carry a causal interpretation \citep{fay2024causal}. In some instances, such estimands offer advantages that make them relevant in social science and biomedical applications. First, the probabilistic index comparing the ranking of outcomes under two potential outcome distributions measures the probability that a randomly chosen unit from $F_1$ has a higher-ranked outcome than a unit from $F_0$. This formulation generally exhibits greater robustness to distributional features. That is, unlike mean contrasts, which can be distorted by heavy tails, skewness, or outliers, probabilistic indices require fewer distribution assumptions and are less sensitive to irregularities in the outcome distribution \citep{hauck2000generalized,acion2006probabilistic}. Second, probabilistic indices are well suited to ordinal outcomes, in which the scientific question concerns relative ordering rather than pure numerical differences. As \citet{de2019tutorial} noted, ``\emph{the probabilistic index provides an ordinal answer to an ordinal question}'', and it relies only on the rank comparison rather than the metric scale or magnitude of differences. Third, the win-type estimand is now increasingly used for prioritized composite outcomes, where win-loss comparisons naturally align with clinical priorities across multiple endpoints. Introduced by \citet{Pocock2012}, the win ratio has gained substantial traction in clinical research with composite outcomes, as it offers a principled way to combine endpoints of varying clinical importance while providing an interpretable summary of treatment benefit in terms of the probability of winning versus losing.

	When drawing causal inferences from randomized experiments or observational studies, intermediate variables such as treatment noncompliance \citep{Angrist1996} and death \citep{Zhang2008} often have important implications on the definition of a causal effect and require careful adjustment. Principal stratification \citep{Frangakis2002} offers one solution by segmenting the population into principal strata based on the joint potential values of the intermediate variable under alternative conditions. The stratum-specific causal effect, referred to as the principal causal effect, can thus be defined and is causally interpretable. However, prior work has predominantly focused on the average causal effect within each principal stratum, and there has been little related development when the generalized causal effect is of interest within each subpopulation. Therefore, we propose a new class of estimands, the \emph{principal generalized causal effect} (PGCE), that accommodates an arbitrary, potentially nonlinear contrast function. As we demonstrate in due course, the PGCE can represent the stratum-specific probabilistic index and the stratum-specific win ratio with an appropriate choice of the contrast function, and it can provide a meaningful assessment of the relative treatment effect within subpopulations. 
	
	Assuming a binary intermediate variable, we study the identification and inference of the PGCE under two structural assumptions, monotonicity \citep{Angrist1996} and principal ignorability \citep{Ding2016}, which are typically invoked for principal stratification analyses. For the principal average causal effect, under monotonicity and principal ignorability, \citet{Ding2016} formalized weighting estimators using principal scores (the conditional probabilities of the latent principal strata given observed covariates); \citet{Jiang2022} provided alternative identification formulas based on the treatment propensity score, the principal score and the potential outcome models, and developed multiply robust and locally efficient estimators; \citet{cheng2023multiply} extended the multiply robust and locally efficient estimators to estimating principal survival causal effect with a right-censored time-to-event outcome; \citet{tong2024doubly} expanded multiply robust and locally efficient estimators to estimate fine principal causal effects with multiple treatments. Despite these advancements, the identification and the semiparametric efficiency theory for the PGCE in the presence of a binary intermediate variable remain largely unclear. 

	The contributions of this article are several-fold. First, we propose a class of PGCE estimands in the presence of a binary intermediate variable and develop nonparametric identification formulas under the assumptions of treatment unconfoundedness, monotonicity, and principal ignorability. Second, we provide a formal treatment of semiparametric efficiency theory for estimating the PGCE estimands and derive their corresponding \emph{efficient influence functions} \citep[EIFs,][]{Bickle1993}. Specifically, we show that when the contrast function is $h(u,v)=u-v$, our EIFs reduce to those in \citet{Jiang2022} as a special case. More broadly, our framework accommodates pairwise comparison estimands with arbitrary contrast functions, offering a substantially more general treatment and establishing the present work as a standalone contribution. A significant complexity of nonlinear contrast functions, however, is that the estimating equations based on the EIFs require nuisance functions defined on pairs of observations, thereby necessitating the theory of U-statistics to construct feasible estimators. We show that the proposed U-statistics are triply robust in the sense that they are consistent and asymptotically normal when two of the three nuisance functions are correctly specified. When all nuisance functions are correctly specified, our estimators are locally efficient. Finally, the construction of the multiply robust U-statistics facilitates the integration of nonparametric estimators for the nuisance functions. To achieve full asymptotic efficiency, we estimate each nuisance function using data-adaptive machine learners under the cross-fitting scheme \citep{Chernozhukov2018}. We discuss the modifications needed for cross-fitting when a nonlinear contrast function is used to define the PGCE, and prove the nonparametric efficiency properties of the debiased machine learning estimators under mild regularity conditions. Example \texttt{R} codes are available at \url{https://github.com/erxc/cusps}.

	\section{Notation, estimands and assumptions} \label{sec:not-asp}

	We consider a study, either experimental or observational, with $n$ individuals and two treatment conditions. We let $Z_i$ denote the treatment assignment of individual $i\in\{1,\ldots,n\}$; $Z_i=1$ indicates the treatment condition and $0$ the control. We use $D_i(z)$ and $Y_i(z)$ to denote the potential outcomes of the intermediate variable and the final outcome variable of interest for participant $i$ under treatment $Z_i=z$ ($z\in\{0,1\}$), respectively. The intermediate variable $D_i\in\{0,1\}$ is assumed to be binary, but there is no restriction on the type of variable for $Y_i$. Under the Stable Unit Treatment Value Assumption (SUTVA), the observed intermediate and final outcomes can be represented as $D_i=Z_iD_i(Z_i)+(1-Z_i)D(1-Z_i)$ and $Y_i=Z_iY_i(Z_i)+(1-Z_i)Y(1-Z_i)$. Let $\bfX_i$ represent the vector of observed pre-treatment covariates, and the observed data for participant $i$ is thus denoted by $\calO_i=\{\bfX_i,Z_i,D_i,Y_i\}$, where $\calO_i$'s are assumed to be independent and identically distributed (i.i.d.). 
	
	\subsection{Principal generalized causal effects}
	
	We operate within the principal stratification framework \citep{Frangakis2002} and define distinct principal strata based on the joint potential values of the intermediate variable under alternative conditions. Specifically, we use $S=\{D(1),D(0)\}$ to denote the principal stratum variable, which has four potential values: $\{10,01,00,11\}$. A key insight under the principal stratification framework is that although $D$ is a post-randomization variable, the stratum variable $S$ is unaffected by the assignment and can be viewed as a pre-treatment variable. Therefore, based on any contrast function $h$, we define the principal generalized causal effect (PGCE) estimand as
	\begin{align} \label{eq:def-pce}
		\tau_h^s=\bbE[h\{Y_1(1),Y_2(0)\}|S_1=S_2=s],~\text{for}~s\in\{10,01,00,11\},
	\end{align}
	where $Y_1(1)$ and $Y_2(0)$ denote the potential outcomes for two randomly selected units from the subpopulation with strata membership $s$. When a linear contrast function is specified as $h(u,v)=u-v$, the PGCE in \eqref{eq:def-pce} reduces to $\tau_h^s=\bbE\{Y(1)-Y(0)|S=s\}$, which is precisely the principal average causal effect studied in the existing principal stratification literature \citep{Frangakis2002}. 
	
	\begin{remark}\label{rmk:estimand}
		One may wonder whether quantities of the form $\tau^{s_1,s_2}_h = \bbE[h\{Y_1(1),Y_2(0)\}|S_1=s_1,S_2=s_2]$ with $s_1 \neq s_2$ can also be considered. While such cross-stratum estimands are in principle identifiable under structural assumptions, they lack a meaningful causal interpretation. Following the definition of causal effects in \citet{Frangakis2002}, the comparison of potential outcomes must be made within a well-defined target population. If $s_1 \neq s_2$, then the two potential outcomes $Y_1(1)$ and $Y_2(0)$ are now drawn from distinct subpopulations characterized by different joint potential intermediate outcomes. In this case, $\tau^{s_1,s_2}_h$ represents a cross-world comparison across heterogeneous subgroups rather than a treatment effect within a common subpopulation, and thus does not conform to the spirit of the original principal stratification framework regarding subpopulation-specific causal effects \citep{Frangakis2002}. As an example, when $D$ denotes the survival status in the truncation-by-death setup (Example \ref{exam-PIM}), comparing $Y_1(1)$ among the protected (those who survive only under treatment) with $Y_2(0)$ among the always-survivors (those who would survive regardless of assignment) conflates two sources of difference: one due to the treatment condition and another due to inherent heterogeneity across subpopulations. This contrast, therefore, cannot be solely attributed to treatment and does not carry a straightforward causal interpretation. For this reason, we restrict attention to $\tau_h^s$ with $s_1 = s_2 = s$.
	\end{remark}
	
	Given $S_1=S_2=s$, we emphasize that the contrast function $h$ is evaluated on two independent replicates drawn from the same principal stratum. This pairwise construction in the causal estimand also motivates the definition of the estimable pairwise mean function in Section \ref{subsec:np-id}, which serves as the building block for our identification strategy and U-statistic-based estimators. To illustrate the PGCE estimand, we next present two examples, while leaving the technical details on constructing U-statistic-based estimators to Section \ref{sec:np-id}. Importantly, when a nonlinear contrast function is used to define the PGCE, $\tau_h^s$ is no longer a simple average of $Y(z)$ within each principal stratum.
	
	\begin{example}[\emph{Probabilistic index under truncation by death}]\label{exam-PIM} 
		When $D$ represents the survival status (alive if $D=1$) that is ascertained before the measurement of the final outcome $Y$ (such as quality of life, which is only well-defined when $D=1$), there exist four principal strata based on the potential survival status. That is, $s=11$ represents the always-survivors who would survive until the measurement of the final outcome regardless of the assignment; $s=10$ and $s=01$ represent the protected and harmed, who would survive until the measurement of the final outcome only under the treatment condition and control condition, respectively; $s=00$ represents the never-survivors who would not survive until the measurement of the final outcome regardless of the assignment. Because the final outcome is typically only well-defined when $D=1$, the survivor average causal effect (SACE) has been suggested as a relevant target of inference with $\mathrm{SACE}=\bbE\{Y(1)-Y(0)|S=11\}$ \citep{Zhang2008, Chen2024}. In clinical research, \citet{acion2006probabilistic} argued that a more intuitive effect size can be measured by the probabilistic index defined with a contrast function $h(u,v)=\bbone(u\geq v)$. Based on this contrast function, we can define the survivor probabilistic index (SPI) as 
		\begin{align} \label{eq:SPI}
			\mathrm{SPI}=\bbP\{Y_1(1)\geq Y_2(0)|S_1=S_2=11\}.
		\end{align}
		Compared with the SACE, the SPI is less sensitive to skewness in the distributions of potential outcomes, heteroskedasticity, and outliers. In the special case of normality and under the equal variance assumption, the SPI is a strictly monotone function of the SACE because $\mathrm{SPI}=\Phi(\mathrm{SACE}/\sqrt{2}\sigma)$, where $\Phi$ is the standard Gaussian cumulative distribution function and $\sigma^2$ is the constant variance of the potential outcomes \citep{wolfe1971constructing}.
	\end{example}
	
	\begin{example}[\emph{Win estimands under noncompliance}]\label{ex-win}
		In experimental studies with ordinal outcomes, \citet{wang2016win} introduced the win ratio estimand for pairwise comparisons, $\bbP\{Y_1(1)>Y_2(0)\}/\bbP\{Y_1(1)<Y_2(0)\}$. This estimand corresponds to a two-dimensional contrast function specified as $h(u,v)=\{\bbone(u>v),\bbone(u<v)\}$. If $D$ represents the actual treatment received that can differ from the assignment, then $S\in\{10,01,00,11\}$ refers to always-takers, compliers, defiers, and never-takers, respectively. Therefore, the win ratio estimands can be extended to define the \emph{compliers win ratio} (CWR) as
		\begin{align} \label{eq:CWR}
			\mathrm{CWR}=\frac{\bbP\{Y_1(1)>Y_2(0)|S_1=S_2=10\}}{\bbP\{Y_1(1)<Y_2(0)|S_1=S_2=10\}}.
		\end{align}
		The CWR represents the odds that a randomly selected complier under the treatment condition has a more favorable outcome than under the control condition. Alternatively, the complier win difference (CWD) or complier net benefit estimand is
		\begin{align}
			\mathrm{CWD}=\bbP\{Y_1(1)>Y_2(0)|S_1=S_2=10\}-\bbP\{Y_1(1)<Y_2(0)|S_1=S_2=10\}.
		\end{align}
		In the context of noncompliance, \citet{Mao2024} defined a local Mann-Whitney treatment effect estimand given by $\bbP\{Y_1(1)\geq Y_2(0)|S_1=S_2=10\}$, which is also a special case of the PGCE. \citet{Mao2024} employed an instrumental variable approach and discussed the point and interval identification of this estimand under monotonicity and exclusion restriction.
	\end{example}

	\subsection{Identification assumptions}
	
	To identify the PGCE from the observed data, we require the following assumptions.
	\begin{assumption}[\emph{Treatment ignorability}] \label{asp:rand}
		$Z_i \independent \{D_i(1),D_i(0),Y_i(1),Y_i(0)\}|\bfX_i$.
	\end{assumption}
	Assumption \ref{asp:rand} rules out confounding between the treatment assignment and intermediate variable as well as final outcomes conditional on observed pre-treatment covariates. This assumption holds by design in randomized experiments. For observational studies, Assumption \ref{asp:rand} can be satisfied if all confounders (common causes) affecting the treatment assignment and the intermediate variable, and those affecting the treatment assignment and the final outcome are measured and included in $\bfX$.
	\begin{assumption}[\emph{Monotonicity}] \label{asp:mono}
		$D_i(1)\geq D_i(0)$ for all $i$.
	\end{assumption}
	Assumption \ref{asp:mono} states that the treatment assignment only affects the intermediate outcome in a non-negative (or non-positive if assume $D_i(1)\leq D_i(0)$) direction and restricts the number of possible strata by excluding the harmed or defiers stratum $S=01$ (or $S=10$). For example, when $D$ represents the survival status, monotonicity holds if the treatment is not expected to harm the survival of each unit. When $D$ represents actual treatment received in noncompliance problems, monotonicity holds if units in the control condition have no access to the active treatment. 
	\begin{assumption}[\emph{Pairwise mean principal ignorability}] \label{asp:pi}
		Given the same values of pre-treatment covariates for a pair of units ($\bfX_1=\bfx_1$ and $\bfX_2=\bfx_2$), the expected contrast function is exchangeable across certain pairs of principal strata. That is,
		\begin{align}
			\begin{split}\label{eq:pi-equation}
				&\bbE[h\{Y_1(1),Y_2(0)\}|S_1=s_1,S_2=s_2,\bfX_1=\bfx_1,\bfX_2=\bfx_2]\\ 
				=~&\bbE[h\{Y_1(1),Y_2(0)\}|S_1=s_1^\prime,S_2=s_2^\prime,\bfX_1=\bfx_1,\bfX_2=\bfx_2] 
			\end{split}
		\end{align}
		if $(s_1,s_2)$ and $(s_1^\prime,s_2^\prime)$ belong to the same set defined by one of the following: (a) $\mathcal{S}_{10}=\{(10,10),(10,00),(11,10),(11,00)\}$; (b) $\mathcal{S}_{11}=\{(10,11),(11,11)\}$; (c) $\mathcal{S}_{00}=\{(00,10),(00,\allowbreak00)\}$; and (d) $\mathcal{S}_{01}=\{(00,11)\}$.
	\end{assumption}
	
	Assumption \ref{asp:pi} rules out unmeasured confounding between principal stratum membership and the final outcome, thereby ensuring that the PGCE estimands can be identified from the observed data. Under Assumptions \ref{asp:rand} - \ref{asp:pi}, for instance, the conditional expectations within group $\mathcal{S}_{10}$ all reduce to $\bbE[h\{Y_1(1),Y_2(0)\}| Z_1=1,D_1=1,Z_2=0,D_2=0,\bfX_1,\bfX_2]$. In other words, within the observed stratum $(Z_1=1,D_1=1,Z_2=0,D_2=0)$, conditional on pre-treatment covariates, the various combinations of possible latent strata including $(S_1=10,S_2=10)$, $(S_1=11,S_2=10)$, $(S_1=10,S_2=00)$, and $(S_1=11,S_2=00)$ are exchangeable with respect to the expected contrast. Similarly, the conditional expected contrasts within $\mathcal{S}_{11}$, $\mathcal{S}_{00}$, and $\mathcal{S}_{01}$ are all equal to the estimable conditional expectations, $\bbE[h\{Y_1(1),Y_2(0)\}|Z_1=1,D_1=1,Z_2=0,D_2=1,\bfX_1,\bfX_2]$, $\bbE[h\{Y_1(1),Y_2(0)\}|Z_1=1,D_1=0,Z_2=0,D_2=0,\bfX_1,\bfX_2]$, and $\bbE[h\{Y_1(1),Y_2(0)\}|Z_1=1,D_1=0,Z_2=0,D_2=1,\bfX_1,\bfX_2]$, respectively (see Table \ref{tab:mapping}(a) for a mapping between the observed data strata and the latent principal strata). That is, with Assumption \ref{asp:pi}, each one of the observed subpopulations, $(Z_1=1,D_1=1,Z_2=0,D_2=0)$, $(Z_1=1,D_1=1,Z_2=0,D_2=1)$, $(Z_1=1,D_1=0,Z_2=0,D_2=0)$, and $(Z_1=1,D_1=0,Z_2=0,D_2=1)$, can be viewed as mixtures of its corresponding subgroups defined by observed covariates only (Table \ref{tab:mapping}(b)).
	
	Assumption \ref{asp:pi} is a generalization of the unitwise mean principal ignorability assumption in \citet{Jiang2022}, since it reduces to that form in the linear contrast case $h(u,v)=u-v$. For nonlinear contrasts, however, expectations cannot be interchanged within the contrast function, and thus there is no clear nesting relationship between the pairwise and unitwise versions of mean principal ignorability. Consider the win estimands under noncompliance (Example \ref{ex-win}) with the contrast function $h(u,v)=\bbone(u>v)$. Applying \eqref{eq:pi-equation} to the first two pairs in $\mathcal{S}_{10}$ yields $\bbP\{Y_1(1) > Y_2(0) | S_1 = 10, S_2 = 10, \bfX_1 = \bfx_1, \bfX_2 = \bfx_2\} = \bbP\{Y_1(1) > Y_2(0) | S_1 = 11, S_2 = 10, \bfX_1 = \bfx_1, \bfX_2 = \bfx_2\}$. This assumption states that, conditional on the same pairwise covariate profiles, the probability that a treated individual has a more favorable outcome than a control individual is the same whether the pair consists of two compliers or of one complier and one always-taker. In other words, there do not exist additional unmeasured factors influencing the conditional probability of winning among pairs in $\mathcal{S}_{10}$. This is neither stronger nor weaker than the unitwise mean principal ignorability assumption, which requires, for example, $\bbE\{Y_1(1) | S_1 = 10, \bfX_1 = \bfx_1\} = \bbE\{Y_1(1) | S_1 = 11, \bfX_1 = \bfx_1\}$.

	\begin{table}[ht]
		\centering
		\caption{Correspondence between latent principal strata pairs, Assumption \ref{asp:pi}, and observed strata pairs defined by assignment and observed intermediate variable.}\label{tab:mapping}
		\begin{subtable}{\textwidth}
			\centering
			\caption{\emph{The mapping from latent principal strata pairs to the conditional pairwise mean defined in observed strata, through Assumption \ref{asp:pi}. There are nine latent strata pairs under monotonicity.}}
			\begin{tabular}{cccc}
				\toprule
				\multicolumn{2}{c}{Latent strata pairs} & Conditions in & Pairwise mean defined in observed strata\\
				$S_1$ & $S_2$ & Assumption \ref{asp:pi} & 
				$\bbE[h\{Y_1(1),Y_2(0)\}| Z_1,D_1,Z_2,D_2,\bfX_1,\bfX_2]$ \\ 
				\midrule
				10 & 10 & (a) & $\{Z_1=1,D_1=1\}$ \& $\{Z_2=0,D_2=0\}$ \\
				10 & 00 & (a) & $\{Z_1=1,D_1=1\}$ \& $\{Z_2=0,D_2=0\}$ \\
				11 & 10 & (a) & $\{Z_1=1,D_1=1\}$ \& $\{Z_2=0,D_2=0\}$ \\
				11 & 00 & (a) & $\{Z_1=1,D_1=1\}$ \& $\{Z_2=0,D_2=0\}$ \\
				10 & 11 & (b) & $\{Z_1=1,D_1=1\}$ \& $\{Z_2=0,D_2=1\}$ \\
				11 & 11 & (b) & $\{Z_1=1,D_1=1\}$ \& $\{Z_2=0,D_2=1\}$ \\
				00 & 10 & (c) & $\{Z_1=1,D_1=0\}$ \& $\{Z_2=0,D_2=0\}$ \\
				00 & 00 & (c) & $\{Z_1=1,D_1=0\}$ \& $\{Z_2=0,D_2=0\}$ \\
				00 & 11 & (d) & $\{Z_1=1,D_1=0\}$ \& $\{Z_2=0,D_2=1\}$ \\
				\bottomrule
			\end{tabular}
		\end{subtable}
		
		\vspace{1em} 
		
		\begin{subtable}{\textwidth}
			\centering
			\caption{\emph{The mapping from observed strata pairs (defined by assignment and observed intermediate variable) as a mixture of latent principal strata pairs, through Assumption \ref{asp:pi}. As the estimand is constructed by contrasting $Y_1(1)$ with $Y_2(0)$, only four observed strata pairs that satisfy $Z_1=1$ and $Z_2=0$ are informative for identification.}}
			\begin{tabular}{cccc}
				\toprule
				\multicolumn{2}{c}{Observed strata pairs} & Conditions in & Mixture components of \\
				Unit 1 & Unit 2 & Assumption \ref{asp:pi} & 
				latent strata pairs\\ 
				\midrule
				$\{Z_1=1,D_1=1\}$ & $\{Z_2=0,D_2=0\}$ & (a) & $(S_1,S_2)\in \mathcal{S}_{10}$\\
				$\{Z_1=1,D_1=1\}$ & $\{Z_2=0,D_2=1\}$ & (b) & $(S_1,S_2)\in \mathcal{S}_{11}$ \\
				$\{Z_1=1,D_1=0\}$ & $\{Z_2=0,D_2=0\}$ & (c) & $(S_1,S_2)\in \mathcal{S}_{00}$ \\
				$\{Z_1=1,D_1=0\}$ & $\{Z_2=0,D_2=1\}$ & (d) & $(S_1,S_2)\in \mathcal{S}_{01}$ \\
				\bottomrule
			\end{tabular}
		\end{subtable}
		
	\end{table}

	Finally, we distinguish Assumptions \ref{asp:rand} and \ref{asp:mono} from Assumption~\ref{asp:pi}. The first two are stated at the individual level, but since the observed data are assumed to be i.i.d., they extend naturally to pairs. For example, individual-level treatment ignorability implies that $(Z_1,Z_2)$ is jointly independent of the potential outcomes and intermediate variables given $(\bfX_1,\bfX_2)$. By contrast, Assumption \ref{asp:pi} must be formulated at the pair level in its minimally sufficient form, because identification with nonlinear contrasts requires linking expectations across observed and latent strata pairs as in Table \ref{tab:mapping}. Assumptions \ref{asp:mono} and \ref{asp:pi} are not testable from the observed data alone. We therefore complement our main results with sensitivity analyses under violations of these assumptions, provided in Section 4.4 and Section S7 of the Online Supplement.

	\section{Nonparametric identification} \label{sec:np-id}

	\subsection{Identification of the PGCEs} \label{subsec:np-id}
	
	To obtain nonparametric identification formulas for the PGCEs, we start by defining components of the observed data distribution. We consider both observational and experimental studies and define the treatment propensity score as $\pi(\bfX)=\bbP(Z=1|\bfX)$, the probability that a participant receives the treatment conditional on pre-treatment covariates \citep{Rosenbaum1983}. In addition, the principal score, $e_s(\bfX)=\bbP(S=s|\bfX)$, is defined as the conditional probability that a participant belongs to the principal stratum $s$ given the covariates \citep{Ding2016}, for $s\in\{10,00,11\}$ under Assumption \ref{asp:mono}. Furthermore, we define $\pi=\bbE\{\pi(\bfX)\}=\bbP(Z=1)$ and $e_s=\bbE\{e_s(\bfX)\}=\bbP(S=s)$. Lastly, we define the pairwise outcome mean within the observed stratum $(Z_1=z_1,D_1=d_1,Z_2=z_2,D_2=d_2)$, as $\mu_{z_1d_1z_2d_2}(\bfX_1,\bfX_2)=\bbE\{h(Y_1,Y_2)|Z_1=z_1,D_1=d_1,Z_2=z_2,D_2=d_2,\bfX_1,\bfX_2\}$, for $z_1,z_2,d_1,d_2\in\{0,1\}$. Below, we provide two examples of the pairwise outcome mean models for specific contrast functions.
	
	\begin{example}[\emph{Pairwise mean function for probabilistic index}] \label{example-mwp}
		To study the probabilistic index estimand, the contrast function is specified as $h(u,v)=\bbone(u\geq v)$. If the potential outcomes are continuous, a Gaussian regression model can be considered such that $Y(z)|D,\bfX \sim \calN(f_z(D,\bfX),\sigma^2)$. In this case, one can derive the pairwise outcome mean as
		\begin{align} \label{eq:mwp-mean}
			\mu_{z_1d_1z_2d_2}(\bfX_1,\bfX_2)=\Phi\left\{\frac{f_{z_1}(d_1,\bfX_1)-f_{z_2}(d_2,\bfX_2)}{\sqrt{2}\sigma}\right\}.
		\end{align}
	\end{example}
	
	\begin{example}[\emph{Pairwise outcome mean for win estimands}] \label{example-wr}
		Consider the win estimands developed in Example \ref{ex-win}, where the contrast function is given by the vector $h(u,v)=\{\bbone(u>v),\bbone(u<v)\}$. For ordinal outcomes, if we specify the proportional odds or ordinal logistic regression model for $Y(z)$ conditional on $D$ and $\bfX$, \citet{Mao2018} has shown that the resulting pairwise outcome mean functions are given by the two-dimensional vector
		\begin{equation}
			\begin{aligned}
				&\mu_{z_1d_1z_2d_2}(\bfX_1,\bfX_2)\\
				&\quad=\left\{\sum_{q=2}^Q\sum_{q'=1}^{q-1}p_{z_1q}(d_1,\bfX_1)p_{z_2q'}(d_2,\bfX_2),\sum_{q=1}^{Q-1}\sum_{q'=q+1}^Qp_{z_1q}(d_1,\bfX_1)p_{z_2q'}(d_2,\bfX_2)\right\},
			\end{aligned}
		\end{equation}
		where $p_{zq}(D,\bfX)$ is the conditional probability of $Y(z)=q\in\{1,\ldots,Q\}$ given $D$ and $\bfX$, and $Q(\geq2)$ is the largest category of the ordinal outcome. 
	\end{example}
	
	For estimating the principal scores, we first define the conditional probability, $p_z(\bfX)=\bbP(D=1|Z=z,\bfX)$, with population fraction $p_z=\bbE\{p_z(\bfX)\}=\bbP(D=1|Z=z)$. Then, under Assumption \ref{asp:mono}, the principal scores can be expressed as $e_{10}(\bfX)=p_1(\bfX)-p_0(\bfX)$, $e_{00}(\bfX)=1-p_1(\bfX)$, and $e_{11}(\bfX)=p_0(\bfX)$. The population fractions of the three principal strata are $e_{10}=p_1-p_0$, $e_{00}=1-p_1$, and $e_{11}=p_0$. We will refer to $\{p_0(\bfX),p_1(\bfX)\}$ as principal scores due to the one-to-one mapping to $\{e_{10}(\bfX),e_{00}(\bfX),e_{11}(\bfX)\}$. Based on these elements, we establish three alternative nonparametric identification formulas for each PGCE estimand, given in Theorem S2 of the Online Supplement. Each formula requires specifying two of the three models---the treatment propensity score, the principal score, and the pairwise outcome mean---rather than all three simultaneously. Furthermore, Theorem S1 of the Online Supplement presents the associated covariate balancing conditions. These balancing conditions provide the foundation for model checking of the treatment propensity and principal score models without peeking at the outcome data.

	\subsection{Estimators based on nonparametric identification formulas} \label{subsec:np-est}
	
	For point estimation, a common approach is to specify parametric regression models for nuisance functions, i.e., $\pi(\bfX;\alpha)$, $p_z(\bfX;\beta)$ or $e_s(\bfX;\beta)$, and $\mu_{z_1d_1z_2d_2}(\bfX_1,\bfX_2;\gamma)$, where $\alpha$, $\beta$, and $\gamma$ are finite-dimensional nuisance parameters, with their corresponding true values denoted by $\alpha^*$, $\beta^*$, and $\gamma^*$. We write the maximum likelihood estimators for these finite-dimensional parameters as $\widehat{\alpha}$, $\widehat{\beta}$, and $\widehat{\gamma}$. To aid further discussions of the statistical properties of estimators for the PGCEs, we introduce notation $\calM_\pi$, $\calM_p$, and $\calM_\mu$ to denote the conditions under which the treatment propensity, principal score, and outcome mean models are correctly specified, respectively. That is, under $\calM_\pi$, we have $\pi(\bfX;\alpha^*)=\pi(\bfX)$; under $\calM_p$, we have $p_z(\bfX;\beta^*)=p_z(\bfX)$ and $e_s(\bfX;\beta^*)=e_s(\bfX)$ for $z\in\{0,1\}$ and $s\in\{10,00,11\}$; and under $\calM_\mu$, we have $\mu_{z_1d_1z_2d_2}(\bfX_1,\bfX_2;\gamma^*)=\mu_{z_1d_1z_2d_2}(\bfX_1,\bfX_2)$ for $z_1,z_2,d_1,d_2\in\{0,1\}$. Additionally, we use the logic notation `+' to denote conditions where more than one nuisance function is correctly specified, and `$\cup$' to denote the condition where at least one nuisance function is correctly specified.
	
	We can obtain plug-in estimators by replacing components in the identification formulas with their empirical counterparts and expectations with empirical averages. For convenience, we first introduce the following notations: $\bbE_nf(\calO)=n^{-1}\sum_{i=1}^nf(\calO_i)$ for any $f(\calO)$, and $\bbU_n f(\calO_1,\calO_2)=\binom{n}{2}^{-1}\sum_{i<j}f(\calO_i,\calO_j)$ for any $f(\calO_1,\calO_2)$. Define the following quantity for any function $f(Y,D,\bfX)$:
	\begin{equation} \label{eq:psi}
		\begin{aligned}
			\psi_{f(Y_z,D_z,\bfX)}(\calO)&=\frac{\bbone(Z=z)[f(Y,D,\bfX)-\bbE\{f(Y,D,\bfX)|\bfX,Z=z\}]}{\bbP(Z=z|\bfX)}\\
			&\qquad+\bbE\{f(Y,D,\bfX)|\bfX,Z=z\}
		\end{aligned}.
	\end{equation}
	One example of \eqref{eq:psi} is $f(Y_z,D_z,\bfX)=D_z$, in which case we obtain the following function:
	\begin{align} \label{eq:psi-dz}
		\psi_{D_z}(\calO)=\frac{\bbone(Z=z)\{D-p_z(\bfX)\}}{\bbP(Z=z|\bfX)} + p_z(\bfX).
	\end{align}
	Equation \eqref{eq:psi-dz} is useful for the estimation of $p_z$, as it leads to a doubly robust estimator once we treat the intermediate variable as the outcome:
	\begin{align} \label{eq:pz-est}
		\widehat{p}_z\equiv\bbE_n\psi_{D_z}(\calO;\widehat{\alpha},\widehat{\beta})=\bbE_n\left[\frac{\bbone(Z=z)\{D-p_z(\bfX;\widehat{\beta})\}}{\pi(X;\widehat{\alpha})}+p_z(\bfX;\widehat{\beta})\right];
	\end{align}
	this estimator improves upon the plug-in estimator $\bbE_n p_z(\bfX;\widehat{\beta})$ in terms of robustness and efficiency. More precisely, it is consistent for $p_z$ under $\calM_\pi\cup\calM_p$. 
	
	For any general contrast function $h$, the plug-in estimators for the PGCEs take the form of U-statistics \citep[\S 12]{vandervaart1998}. For example, the identification results in Theorem S2 of the Online Supplement motivate the following estimators for $\tau_h^{10}$:
	\begin{align*}
		\widehat{\tau}_{h,\pi+p}^{10}&=\bbU_n\left[\left\{\frac{e_{10}(\bfX_1;\widehat{\beta})}{\widehat{p}_1-\widehat{p}_0}\frac{D_1}{p_1(\bfX_1;\widehat{\beta})}\frac{Z_1}{\pi(\bfX_1;\widehat{\alpha})}\right\}\left\{\frac{e_{10}(\bfX_2;\widehat{\beta})}{\widehat{p}_1-\widehat{p}_0}\frac{1-D_2}{1-p_0(\bfX_2;\widehat{\beta})}\frac{1-Z_2}{1-\pi(\bfX_2;\widehat{\alpha})}\right\}h(Y_1,Y_2)\right], \displaybreak[0]\\
		\widehat{\tau}_{h,\pi+\mu}^{10}&=\bbU_n\left(\left[\frac{D_1Z_1/\pi(\bfX_1;\widehat{\alpha})-D_1(1-Z_1)/\{1-\pi(\bfX_1;\widehat{\alpha})\}}{\widehat{p}_1-\widehat{p}_0}\right]\right.\displaybreak[0]\\
		&\qquad\qquad\times\left.\left[\frac{D_2Z_2/\pi(\bfX_2;\widehat{\alpha})-D_2(1-Z_2)/\{1-\pi(\bfX_2;\widehat{\alpha})\}}{\widehat{p}_1-\widehat{p}_0}\right]\mu_{1100}(\bfX_1,\bfX_2;\widehat{\gamma})\right),\displaybreak[0]\\
		\widehat{\tau}_{h,p+\mu}^{10}&=\bbU_n\left[\left\{\frac{p_1(\bfX_1;\widehat{\beta})-p_0(\bfX_1;\widehat{\beta})}{\widehat{p}_1-\widehat{p}_0}\right\}\left\{\frac{p_1(\bfX_2;\widehat{\beta})-p_0(\bfX_2;\widehat{\beta})}{\widehat{p}_1-\widehat{p}_0}\right\}\mu_{1100}(\bfX_1,\bfX_2;\widehat{\gamma})\right].
	\end{align*}
	For brevity, we will use $\tau_h^{10}$ as a focused example in the remainder of the article. Results for $\tau_h^{00}$ and $\tau_h^{11}$ are analogous and are given in Section S2 of the Online Supplement. These estimators have strong requirements on model specification to estimate the PGCEs consistently. To reiterate, $\widehat{\tau}_{h,\pi+p}^s$ for $s\in\{10,00,11\}$ are consistent under $\calM_{\pi+p}$, that is, when the treatment propensity and the principal score are both correctly specified; $\widehat{\tau}_{h,\pi+\mu}^s$ for $s\in\{10,00,11\}$ are consistent under $\calM_{\pi+\mu}$, since the doubly robust estimator for $p_z$ is consistent under $\calM_\pi\cup\calM_p$; and $\widehat{\tau}_{h,p+\mu}^s$ for $s\in\{10,00,11\}$ are consistent under $\calM_{p+\mu}$.
	
	\section{Multiply robust and efficient U-statistics} \label{sec:eif}
	
	\subsection{The efficient influence function for the PGCE estimand} \label{subsec:eif}
	
	We first develop more efficient estimators with $n^{1/2}$ convergence rate under feasible conditions for the PGCEs by obtaining their corresponding EIFs \citep{Bickle1993}. The EIF is a nonparametric function of observed data that characterizes the target estimand, and its form is instrumental in developing optimal estimators. Several key properties of the EIF suggest its promise for constructing optimal estimators. First, the EIF integrates multiple nuisance models and facilitates estimators that are robust against partial model misspecification. Second, the lower bound on the asymptotic variance is the variance of the EIF, providing a benchmark for defining efficient estimators. Third, the EIF provides a device for incorporating data-adaptive machine learners for nuisance parameters, yielding optimal estimators that achieve the asymptotic variance lower bound with a $n^{1/2}$ convergence rate \citep{Chernozhukov2018}. We explore these properties by first formalizing the EIF for each PGCE. Because the PGCEs are of a ratio structure, we define $\tau_h^s=\tau_{h,N}^s/\tau_D^s$ for $s\in\{10,00,11\}$, where, for example, $\tau_{h,N}^{10}=\bbE\{e_{10}(\bfX_1)e_{10}(\bfX_2)\allowbreak\mu_{1100}(\bfX_1,\bfX_2)\}$ and $\tau_D^{10}=(p_1-p_0)^2$. This suggests that the EIF for $\tau_h^s$, $\varphi_h^s(\calO)$, is of the form:
	\begin{align*}
		\varphi_h^s(\calO) \equiv \frac{1}{\tau_D^s}\varphi_{h,N}^s(\calO) - \frac{\tau_{h,N}^s}{\tau_D^s{}^2}\varphi_D^s(\calO) = \frac{\varphi_{h,N}^s(\calO)-\tau_h^s\varphi_D^s(\calO)}{\tau_D^s},
	\end{align*}
	where $\varphi_{h,N}^s(\calO)$ and $\varphi_{D}^s(\calO)$ are EIFs for $\tau_{h,N}^s$ and $\tau_{D}^s$, respectively. Let $F_{zd}(\cdot|\bfx)$ denote the conditional distribution of $Y_i(z)$ given $(Z_i,D_i,\bfX_i)=(z,d,\bfx)$, and $F_{\bfX}(\cdot)$ denote the marginal distribution of $\bfX$. We define marginalized functions of the observed outcome, $\nu_{1,zd}^s(Y)=\int e_s(\bfx)h(Y,y)\rmd F_{zd}(y|\bfx)\allowbreak\rmd F_{\bfX}(\bfx)$ and $\nu_{0,zd}^s(Y)=\int e_s(\bfx)h(y,Y)\allowbreak\rmd F_{zd}(y|\bfx)\allowbreak\rmd F_{\bfX}(\bfx)$, for $z,d\in\{0,1\}$ and $s\in\{10,00,11\}$. Then, the following Theorem \ref{thm:eif} is obtained.
	\begin{theorem}[\emph{EIF}] \label{thm:eif}
		Under Assumptions \ref{asp:rand} - \ref{asp:pi}, the EIF for $\tau_h^{10}$ is $\varphi_h^{10}(\calO)=\{\varphi_{h,N}^{10}(\calO)-\tau_h^{10}\varphi_{D}^{10}(\calO)\}/(p_1-p_0)^2$, where $\varphi_{h,N}^{10}(\calO)=\phi_h^{10}(\calO)-2\tau_{h,N}^{10}$ and
		\begin{align*}
			\phi_h^{10}(\calO)=&~\frac{ZDe_{10}(\bfX)}{\pi(\bfX)p_1(\bfX)}\left\{\nu_{1,00}^{10}(Y)-m_1^{10}(\bfX)\right\}+\frac{(1-Z)(1-D)e_{10}(\bfX)}{\{1-\pi(\bfX)\}\{1-p_0(\bfX)\}}\left\{\nu_{0,11}^{10}(Y)-m_0^{10}(\bfX)\right\}\displaybreak[0]\\
			&+\{\psi_{D_1}(\calO)-\psi_{D_0}(\calO)\}\left\{m_1^{10}(\bfX)+m_0^{10}(\bfX)\right\},
		\end{align*}
		with $m_1^{10}(\bfX)=\int e_{10}(\bfx)\mu_{1100}(\bfX,\bfx)\rmd F_{\bfX}(\bfx)$, $m_0^{10}(\bfX)=\int e_{10}(\bfx)\mu_{1100}(\bfx,\bfX)\rmd F_{\bfX}(\bfx)$, and
		\begin{align*}
			\varphi_{D}^{10}(\calO)=2\{\psi_{D_1}(\calO)-\psi_{D_0}(\calO)\}(p_1-p_0) - 2\tau_{D}^{10}.
		\end{align*}
	\end{theorem}
	
	For brevity, the EIFs for $\tau_h^{00}$ and $\tau_h^{11}$ are given in Theorem S3 of Online Supplement Section S3.1. The EIFs for the PGCEs are derived following the chain rule via the method of G\^{a}teaux derivatives with respect to the distribution of observed data, $\calO$ \citep{Ichimura2022,Hines2022}.	Based on Theorems \ref{thm:eif} and S3, the asymptotic variances of estimators for the PGCEs are lower bounded by the respective semiparametric efficiency bounds, $\bbE\{\varphi_h^s(\calO)^2\}$ for $s\in\{10,00,11\}$.

	\subsection{Multiply robust estimators} \label{subsec:m-r-est}
	
	The EIFs imply the semiparametric efficiency lower bound and pave the way for the formulation of optimal estimators for the PGCEs. However, each EIF involves unknown marginal functions such as $m_z^s(\cdot)$ and $\nu_{1,00}^{10}(\cdot)$ that are defined as integrals of the pairwise mean function and require care for constructing the final estimator. We use U-statistics to construct the final estimator, which has the same influence function as in Theorem \ref{thm:eif}, thereby achieving local semiparametric efficiency with correctly specified nuisance functions. To this end, we first notice that $\tau_h^{10}$ can be rewritten as
	\begin{align} \label{eq:eif-estimand}
		\tau_h^{10}=\frac{2^{-1}\bbE(\phi_h^{10})}{\{\bbE(\psi_{D_1}-\psi_{D_0})\}^2}.
	\end{align}
	
	Let $\theta$ denote the collection of parametric nuisance function parameters and $\vartheta$ denote a subset of $\theta$, i.e., $\theta=\{\alpha,\beta,\gamma\}$ and $\vartheta=\{\alpha,\beta\}$, with corresponding maximum likelihood estimators $\widehat{\theta}=\{\widehat{\alpha},\widehat{\beta},\widehat{\gamma}\}$ and $\widehat{\vartheta}=\{\widehat{\alpha},\widehat{\beta}\}$, as well as true values $\theta^*=\{\alpha^*,\beta^*,\gamma^*\}$ and $\vartheta^*=\{\alpha^*,\beta^*\}$. Then, the denominator in \eqref{eq:eif-estimand} can be estimated by $[\bbE_n\{\psi_{D_1}(\widehat{\vartheta})-\psi_{D_0}(\widehat{\vartheta})\}]^2$, where $\bbE_n\psi_{D_z}(\widehat{\vartheta})\equiv\bbE_n\psi_{D_z}(\calO;\widehat{\alpha},\widehat{\beta})$ is the doubly robust estimator given in \eqref{eq:pz-est}. Estimating the numerator in \eqref{eq:eif-estimand} is more complicated. To target these terms, we construct the following function of the observed data pairs:
	\begin{align*}
		&g_h^{10}(\calO_i,\calO_j;\theta)\displaybreak[0]\\
		&=2^{-1}\bigg[\frac{Z_iD_ie_{10}(\bfX_i;\beta)}{\pi(\bfX_i;\alpha)p_1(\bfX_i;\beta)}\times\frac{(1-Z_j)(1-D_j)e_{10}(\bfX_j;\beta)}{\{1-\pi(\bfX_j;\alpha)\}\{1-p_0(\bfX_j;\beta)\}}\left\{h(Y_i,Y_j)-\mu_{1100}(\bfX_i,\bfX_j;\gamma)\right\} \displaybreak[0]\\
		&\qquad\qquad+\frac{Z_jD_je_{10}(\bfX_j;\beta)}{\pi(\bfX_j;\alpha)p_1(\bfX_j;\beta)}\times\frac{(1-Z_i)(1-D_i)e_{10}(\bfX_i;\beta)}{\{1-\pi(\bfX_i;\alpha)\}\{1-p_0(\bfX_i;\beta)\}}\left\{h(Y_j,Y_i)-\mu_{1100}(\bfX_j,\bfX_i;\gamma)\right\} \displaybreak[0]\\
		&\qquad\qquad+\{\psi_{D_1}(\calO_i;\alpha,\beta)-\psi_{D_0}(\calO_i;\alpha,\beta)\}\{\psi_{D_1}(\calO_j;\alpha,\beta)-\psi_{D_0}(\calO_j;\alpha,\beta)\}\mu_{1100}(\bfX_i,\bfX_j;\gamma) \displaybreak[0]\\
		&\qquad\qquad+\{\psi_{D_1}(\calO_j;\alpha,\beta)-\psi_{D_0}(\calO_j;\alpha,\beta)\}\{\psi_{D_1}(\calO_i;\alpha,\beta)-\psi_{D_0}(\calO_i;\alpha,\beta)\}\mu_{1100}(\bfX_j,\bfX_i;\gamma)\bigg],
	\end{align*}
	for $1\leq i\neq j\leq n$ (expressions for $g_h^{00}$ and $g_h^{11}$ are given in Section S4.1 of the Online Supplement). For $g_h^s(\calO_i,\calO_j;\theta)$, it can be shown that $2\bbE\{g_h^s(\calO_i,\calO_j;\theta^*)|\calO_i\} = \phi_h^s(\calO_i)$ and $\bbE\{g_h^s(\calO_i,\allowbreak\calO_j;\theta^*)\} = 2^{-1}\bbE\{\phi_h^s(\calO_i)\}=\tau_{h,N}^s$. Thus, the numerators can be estimated using an empirical pairwise average as $\bbU_n g_h^s(\widehat{\theta})\equiv\binom{n}{2}^{-1}\sum_{i<j}g_h^s(\calO_i,\calO_j;\widehat{\theta})$ for $s\in\{10,00,11\}$. Therefore, we obtain the following estimator for $\tau_h^{10}$ expressed in \eqref{eq:eif-estimand}:
	\begin{align} \label{eq:eif-estimator}
		\widehat{\tau}_{h,\tr}^{10}=\frac{\bbU_n g_h^{10}(\widehat{\theta})}{[\bbE_n\{\psi_{D_1}(\widehat{\vartheta})-\psi_{D_0}(\widehat{\vartheta})\}]^2}, 
	\end{align}
	which is essentially a scaled U-statistic. Estimators $\widehat\tau_{h,\tr}^{00}$ and $\widehat\tau_{h,\tr}^{11}$ are similarly constructed and defined in Section S4.1 of the Online Supplement. The asymptotic properties of these estimators are summarized in Theorem \ref{thm:robust}.
	\begin{theorem}[\emph{Multiply robustness and efficiency}]\label{thm:robust}
		Under Assumptions \ref{asp:rand} - \ref{asp:pi}, and suppose there exists some $\varepsilon\in(0,1)$ such that $\varepsilon<\{\pi(\bfX;\alpha^*),\pi(\bfX;\widehat{\alpha})\}<1-\varepsilon$, and $\{p_1(\bfX;\beta^*),\allowbreak p_1(\bfX;\widehat{\beta}), 1-p_0(\bfX;\beta^*),1-p_0(\bfX;\widehat{\beta})\}>\varepsilon$ for all $\bfX\in\bm{\calX}$. Then, each estimator $\widehat{\tau}_{h,\tr}^s$, $s\in\{10,00,11\}$, is (a) triply robust such that it is consistent for $\tau_h^s$ under $\calM_{\pi+p}\cup\calM_{\pi+\mu}\cup\calM_{p+\mu}$; (b) asymptotically linear with influence function $\varphi_h^s$ under $\calM_{\pi+p+\mu}$ such that it achieves the semiparametric efficiency lower bound.
	\end{theorem}
	
	By Theorem \ref{thm:robust}, $\widehat{\tau}_{h,\tr}^s$ is consistent if any two of the three parametric nuisance functions are correctly specified and is semiparametrically efficient if all three nuisance models are correctly specified. This triple robustness property of $\widehat{\tau}_{h,\tr}^s$ is obtained by separately investigating the properties of its numerator and the denominator. To account for the uncertainty in estimating the parametric nuisance functions, we estimate the variances of the triply robust estimators via the nonparametric bootstrap.

	\subsection{Debiased machine learning estimators} \label{subsec:debiased-ml}
	
	We study further improvements to the parametric multiply robust estimator via debiased machine learning, which is similarly based on orthogonal moment functions \citep{Chernozhukov2018} or EIFs, but allows efficient estimation of causal effects using data-adaptive nuisance estimators. Because the Neyman orthogonality holds for the EIF \citep{Chernozhukov2022}, perturbations in the first-step estimates typically have no significant influence on the local properties of the desired estimator for the causal estimand. In addition, the mixed bias property of the EIF representation allows one to obtain $n^{1/2}$-consistent and asymptotically normal estimators when one can estimate a subset of nuisance functions at sufficiently fast rates. Compared with parametric working models, machine learning based working models for the nuisance functions are more robust to potential misspecification. To control the remainder empirical process terms of the debiased machine learning estimators to be asymptotically negligible, a sample-splitting procedure---cross-fitting---is adopted. In this procedure, the entire sample is divided into separate parts for estimating the nuisance functions and the causal estimand, respectively \citep{Chernozhukov2018}. Mild regulatory conditions, mainly the consistency of machine learning first-step estimates, are required for the $n^{1/2}$-consistency and asymptotic normality of the resulting debiased machine learning estimator \citep{kennedy2023semiparametric}.
	
	We adopt the sample-splitting scheme for U-statistics provided in \citet{escanciano2022machine}. Specifically, the sample index set $\{1,\ldots,n\}$ is first divided into $K$ subsets of similar sizes, forming a partition $\calI=\{I_1,\ldots,I_K\}$. Then, upon $\calI$, a partition, $\calP$, can be formed for all pairs $\{(i,j),1\leq i<j\leq n\}$, which has $L=K(K+1)/2$ elements, i.e., $\calP=\{P_1,\ldots,P_L\}$. \citet{Chernozhukov2018} suggests setting $K$ to moderate values, such as 4 or 5, so that the sizes of elements in $\calI$ are $O(n)$. We use the following example to demonstrate the definition of $\calI$ and $\calP$.
	
	\begin{example}[\emph{Sample splitting for U-statistics}]
		Consider a scenario where $n=10$ and $K=3$. We can first form the partition $\calI=\{I_1,I_2,I_3\}$ with $I_1=\{1,2,3\}$, $I_2=\{4,5,6,7\}$, and $I_3=\{8,9,10\}$. Then, the partition $\calP$ is formed as $$\{(I_1,I_1),(I_1,I_2),(I_1,I_3),(I_2,I_2),(I_2,I_3),(I_3,I_3)\}.$$ 
		For $\calP$, elements formed by $I_{k_1}$ and $I_{k_2}$ with $k_1<k_2$ are straightforward Cartesian product of $I_{k_1}$ and $I_{k_2}$. For instance, 
		$$(I_1,I_2)=\{(1,4),(1,5),(1,6),(1,7),(2,4),(2,5),(2,6),(2,7),(3,4),\allowbreak(3,5),(3,6),(3,7)\}.$$ 
		Elements formed by $I_k$ itself, on the other hand, need to exclude pairs consisting of the same indices, e.g., $(I_1,I_1)=\{(1,2),(1,3),(2,3)\}$. According to this sampling-splitting scheme, all $\binom{10}{2}=45$ possible pairs are divided into $L=6$ subsets. 
	\end{example}
	
	With a slight abuse of notation, let $\widehat{\theta}=\{\widehat{\pi}, \widehat{p}_z,\widehat{\mu}_{z_1d_1z_2d_2}\}$ and $\widehat{\vartheta}=\{\widehat{\pi}, \widehat{p}_z\}$ denote machine learning estimators of nuisance functions $\theta=\{\pi, p_z, \mu_{z_1d_1z_2d_2}\}$ and $\vartheta=\{\pi, p_z\}$, respectively, as well as true functions $\theta^*=\{\pi^*, p_z^*, \mu_{z_1d_1z_2d_2}^*\}$ and $\vartheta^*=\{\pi^*, p_z^*\}$. For a given partition $\calP$ (based on the initial partition of singletons $\calI$), we obtain intermediate estimates of the target estimand using observations in each $P_l$ (or $I_k$) with $\widehat{\theta}_l$ (or $\widehat{\vartheta}_k$) estimated using observations in $P_l^c$ (or $I_k^c$), those are not present in $\widehat{\theta}_l$ (or $\widehat{\vartheta}_k$). By pooling these intermediate estimates together, we obtain the debiased machine learning estimator for $\tau_h^{10}$ as
	\begin{align} \label{eq:eif-estimator-ml}
		\widehat{\tau}_{h,\dml}^{10}=\frac{\bbU_{n,L} g_h^{10}(\widehat{\theta})}{[\bbE_{n,K}\{\psi_{D_1}(\widehat{\vartheta})-\psi_{D_0}(\widehat{\vartheta})\}]^2},
	\end{align}
	where $\bbU_{n,L} g_h^{10}(\widehat{\theta}) = \binom{n}{2}^{-1}\sum_{l=1}^L\sum_{(i,j)\in P_l} g_h^{10}(\calO_i,\calO_j;\widehat{\theta}_l)$ and $\bbE_{n,K}\psi_{D_z}(\widehat{\vartheta}) = n^{-1}\sum_{k=1}^K\sum_{i\in I_k}\allowbreak\psi_{D_z}(\calO_i;\widehat{\vartheta}_k)$.

	The debiased machine learning estimators $\widehat{\tau}_{h,\dml}^{00}$ and $\widehat{\tau}_{h,\dml}^{11}$ can be constructed in a similar fashion and are given in Section S5 of the Online Supplement. To establish the asymptotic properties of $\widehat{\tau}_{h,\dml}^s$, we first assume $\theta$ and $\vartheta$ take values in $\Theta$, a convex subset of some normed vector space, and define $\Theta^*\subset\Theta$ as a nuisance realization set such that, for all $l$ and $k$, $\widehat{\theta}_l$ and $\widehat{\vartheta}_k$ take values in this set with high probability. In addition, we assume true nuisance realizations $\theta^*,\vartheta^*\in\Theta^*$. The following Theorem \ref{thm:ml} summarizes the convergence and efficiency properties of the debiased machine learning estimators.
	\begin{theorem}[\emph{Nonparametric efficiency}] \label{thm:ml}
		Under Assumptions \ref{asp:rand} - \ref{asp:pi}, and 
		\begin{itemize}
			\item[(a)] $\bbE|g_h^s(\calO_i,\calO_j;\theta^*)|^2<\infty$ and $\bbE|\psi_{D_z}(\calO_i;\vartheta^*)|^2<\infty$;
			\item[(b)] $\widehat{\theta}_l$ is consistent for $\theta$ and $\widehat{\vartheta}_k$ is consistent for $\vartheta$, with $\bbE|g_h^s(\calO_i,\calO_j;\widehat{\theta}_l)-g_h^s(\calO_i,\calO_j;\theta^*)|^2\xrightarrow{p}0$, and $\bbE|\psi_{D_z}(\calO_i;\widehat{\vartheta}_k)-\psi_{D_z}(\calO_i;\vartheta^*)|^2\xrightarrow{p}0$, for all $l$ and $k$;
			\item[(c)] $\|\widehat{\theta}_l-\theta^*\|=o_\bbP(n^{-1/4})$ and $\|\widehat{\vartheta}_k-\vartheta^*\|=o_\bbP(n^{-1/4})$, for all $l$ and $k$, where $\|\cdot\|$ denotes the $\calL_2$-norm;
			\item[(d)] $g_{h,1}^s(\calO_1;\theta)\equiv\bbE g_h^s(\calO_1,\calO_2;\theta)$ and $\psi_{D_z}(\calO;\vartheta)$ are $\bbP$-Donsker with probability approaching 1, and there exists $C>0$ such that $|\bbE g_h^s(\calO_i,\calO_j;\theta)-\tau_{h,N}^{s*}|\leq C\|\theta-\theta^*\|^2$ and $|\bbE\psi_{D_z}(\calO_i;\vartheta)-p_z^*|\leq C\|\vartheta-\vartheta^*\|^2$ for $\theta,\vartheta\in\Theta^*$ when $\|\theta-\theta^*\|$ and $\|\vartheta-\vartheta^*\|$ are small,
		\end{itemize}
		then $\widehat{\tau}_{h,\dml}^s$ is $n^{1/2}$-consistent, asymptotically normal, and has influence function $\varphi_h^s$, thereby achieving the semiparametric efficiency lower bound characterized by the variance of the EIF.
	\end{theorem}
	
	Theorem \ref{thm:ml} establishes the consistency and asymptotic normality of the debiased machine learning estimators, building on the results developed for orthogonal moment functions given in \citet[Lemma 15]{Chernozhukov2022} and \citet[Lemma 4]{escanciano2022machine}. Regulatory conditions in Theorem \ref{thm:ml} are mostly mild. Condition (a) is a standard boundedness requirement on functions $g_h^s(\calO_i,\calO_j;\theta^*)$ and $\psi_{D_z}(\calO_i;\vartheta^*)$. Condition (b) assumes that machine learning first-step estimators are consistent, as well as the $\calL_2$-moment convergence of $g_h^s(\calO_i,\calO_j;\theta)$ and $\psi_{D_z}(\calO_i;\vartheta)$ are attainable, which are generally achievable for existing machine learners. Condition (c) and part of condition (d) together constitute the small bias assumption for $\widehat{\theta}_l$ and $\widehat{\vartheta}_k$; Condition (c), in particular, specifies the required convergence rate for first-step estimates at $n^{-1/4}$, which could be satisfied by most existing data-adaptive machine learning methods \citep{Chernozhukov2018}. The $\bbP$-Donsker class assumption in condition (d) regulates the complexity of the function classes for both the numerator and denominator; similar conditions have been previously invoked in the debiased machine learning literature \citep{Vansteelandt2022,Jiang2022}. Finally, for statistical inference, we estimate the variances of the debiased machine learning estimators given in \eqref{eq:eif-estimator-ml} via the nonparametric bootstrap.

	\subsection{Sensitivity analysis for monotonicity} \label{subsec:sensitivity-analysis}
	
	To point identify PGCEs, a central assumption is monotonicity (Assumption \ref{asp:mono}), which rules out the $01$ stratum. This stratum may be relevant, for example, when comparing two active interventions with unknown survival benefits against non-mortality outcomes. We thus describe a sensitivity analysis strategy to evaluate the robustness of the proposed estimators in the event of departure from monotonicity. Following the approach in \citet{Ding2016}, we define the sensitivity function $\eta(\bfX)$ as the ratio between the proportion of strata $01$ and $10$ conditional on the covariates, i.e., $\eta(\bfX) = \bbP(S=01|\bfX)/\bbP(S=10|\bfX)$, where $\eta(\bfX)\in[0,\infty)$; the monotonicity holds if $\eta(\bfX)=0$, but $\eta(\bfX)>0$ implies the existence of the $01$ stratum. The PCGEs in this scenario are thus $\tau_{h,\eta}^s$ for $s\in\{10,01,00,11\}$. To proceed, we first notice that, for $\eta(\bfX)\neq 1$, the principal scores can be obtained as:
	\begin{equation} \label{eq:ps-sensitivity}
		\begin{aligned}
			&e_{10,\eta}(\bfX)=\frac{p_1(\bfX)-p_0(\bfX)}{1-\eta(\bfX)}, ~~ e_{01,\eta}(\bfX)=\frac{\eta(\bfX)\{p_1(\bfX)-p_0(\bfX)\}}{1-\eta(\bfX)}, \\
			&e_{00,\eta}(\bfX)=1-p_0(\bfX)-\frac{p_1(\bfX)-p_0(\bfX)}{1-\eta(\bfX)}, ~~ e_{11,\eta}(\bfX)=p_1(\bfX)-\frac{p_1(\bfX)-p_0(\bfX)}{1-\eta(\bfX)},
		\end{aligned}
	\end{equation}
	with $0\leq\eta(\bfX)\leq 1-\{p_1(\bfX)-p_0(\bfX)\}/\min\{p_1(\bfX),1-p_0(\bfX)\}$, and the proportions of principal strata by $e_{s,\eta}=\bbE\{e_{s,\eta}(\bfX)\}$ for all $s$. When $\eta(\bfX)=1$, strata $10$ and $01$ have equal proportions, which corresponds to the situation where the treatment has zero average causal effect on $S$, and the principal scores are not identifiable \citep[Section S4.3]{Jiang2022}. Thus, this boundary case was ruled out. The identification of the PGCEs without monotonicity also requires an alternative version of the principal ignorability (Assumption S1), which we provide in Section S7.1 of the Online Supplement. 
	
	With a fixed value of the sensitivity function $\eta(\bfX)$, we derive bias-corrected multiply robust and debiased machine learning estimators for the PGCEs, leveraging the EIFs obtained without assuming monotonicity. The EIFs and the explicit forms of the bias-corrected estimators are given in Section S7.1 of the Online Supplement. Importantly, these multiply robust estimators remain triply robust under $\calM_{\pi+p}\cup\calM_{\pi+\mu}\cup\calM_{p+\mu}$, as summarized in the Theorem S6 in Section S7.1 of the Online Supplement.

	\section{Simulation studies} \label{sec:sim}
	
	\subsection{Probabilistic index with an intermediate variable}\label{sim-PI}
	
	In the first simulation study, we focus on the probabilistic index estimand introduced in Examples \ref{exam-PIM} and \ref{example-mwp}, in which the potential outcomes are continuous and drawn from a normal distribution. We first generate four pre-treatment covariates $\bfX_i=(X_{i1},X_{i2},X_{i3},X_{i4})^\top$, where $X_{i1},X_{i2},X_{i3}\overset{\text{i.i.d.}}{\sim} \calN(0,1)$, and $X_{i4}\overset{\text{i.i.d.}}{\sim} \calB(0.5)$. Then, we draw the treatment assignment from $Z_i|\bfX_i\sim \calB\{\expit(-X_{i1}+0.5X_{i2}-0.25X_{i3}-0.1X_{i4})\}$. Finally, the potential intermediate variable and the potential outcome are drawn from $D_i(Z_i)|Z_i,\bfX_i\sim \calB\{\expit(-1+2Z_i+X_{i1}-0.8X_{i2}+0.6X_{i3}-X_{i4})\}$, and $Y_i(Z_i)|D_i,Z_i,\bfX_i\sim\calN(10+2Z_i-D_i+8X_{i1}+6X_{i2}+9X_{i3}+7X_{i4},1)$ for $i\in\{1,\ldots,n\}$, where, $\expit(x)=1/\{1+\exp(-x)\}$. We consider a binary intermediate variable ($D$ can be considered as the treatment receipt in the context of non-compliance) and assume the interest lies in estimating all three principal probabilistic index estimands ($\tau_h^{10}$, $\tau_h^{00}$ and $\tau_h^{11}$ with $h(u,v)=\bbone(u\geq v)$). In addition, we create transformed covariates, $\widetilde{\bfX}_i=(\widetilde{X}_{i1},\widetilde{X}_{i2},\widetilde{X}_{i3},\widetilde{X}_{i4})^\top$, where
	\begin{align*}
		\widetilde{X}_{i1}=\exp(0.5X_{i1}), ~\widetilde{X}_{i2}=\frac{X_{i2}}{1+X_{i1}},~\widetilde{X}_{i3}=\left(\frac{X_{i2}X_{i3}}{25}+0.6\right)^3,~\widetilde{X}_{i4}=\frac{X_{i3}X_{i4}}{\sqrt{2}},
	\end{align*}
	are used to simulate scenarios where the nuisance functions are misspecified. We fit the model using $\bfX_i$ in scenarios where the respective nuisance function is correctly specified, and $\widetilde{\bfX}_i$ otherwise. It is worth noting that, although in theory, machine learning methods are flexible with richer model spaces as $n$ increases, model misspecification can still be relevant to the debiased machine learning estimators, as their estimation accuracy in finite samples can be affected by the quality of the input covariate matrix \citep{cheng2025inverting}. 
	
\begin{figure}[htbp]
		\centering
		\begin{subfigure}{\textwidth}
			\centering
			\includegraphics[width=\textwidth]{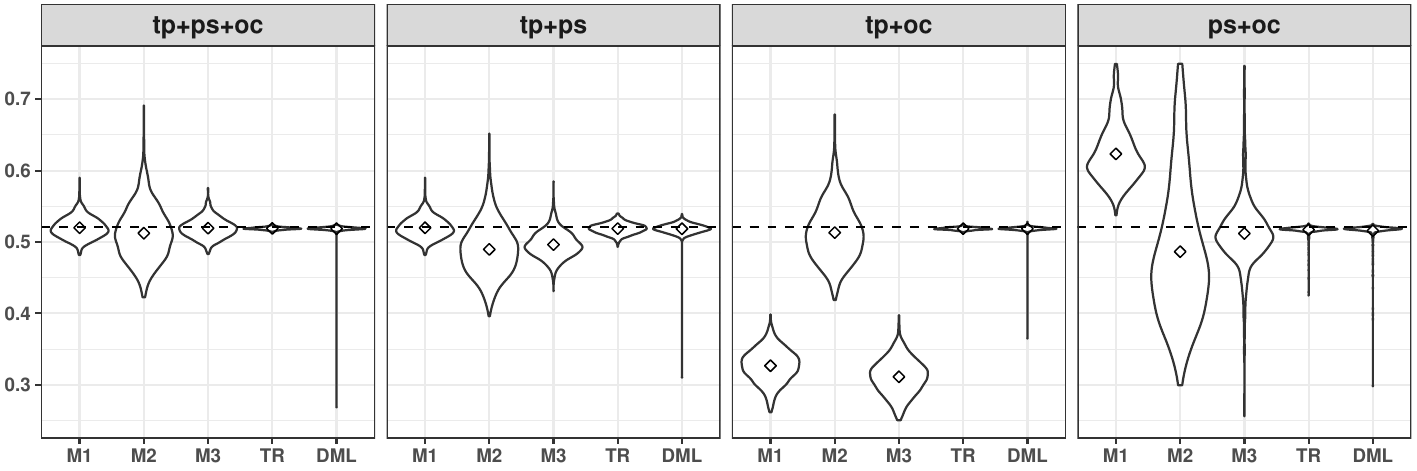}
			\caption{$\tau_h^{10}$}
			\label{subfig:mw-10}
		\end{subfigure}
		\\
		\begin{subfigure}{\textwidth}
			\centering
			\includegraphics[width=\textwidth]{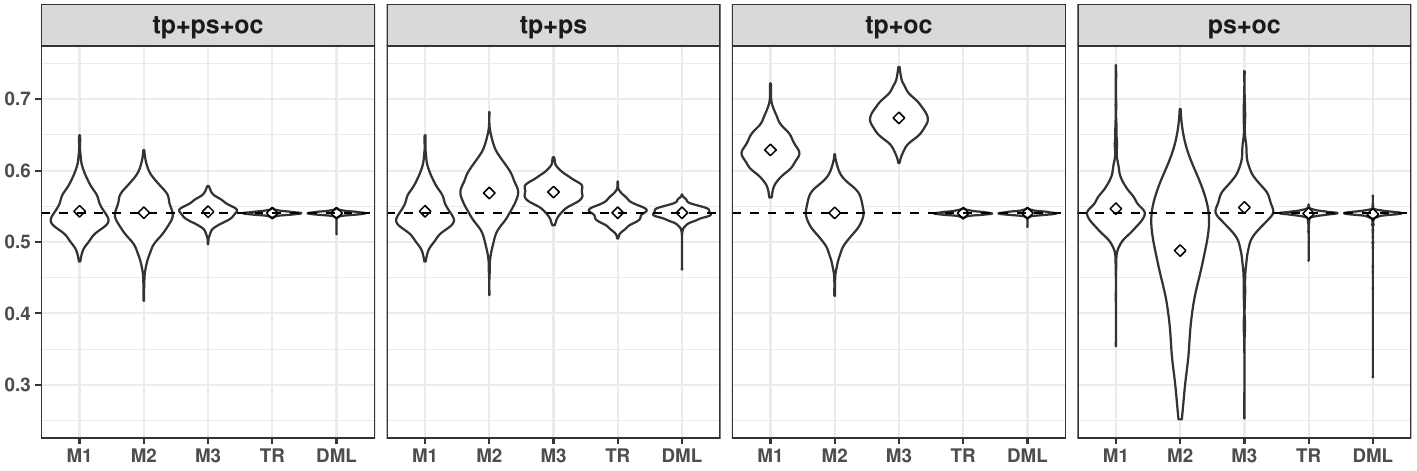}
			\caption{$\tau_h^{00}$}
			\label{subfig:mw-00}
		\end{subfigure}
		\\
		\begin{subfigure}{\textwidth}
			\centering
			\includegraphics[width=\textwidth]{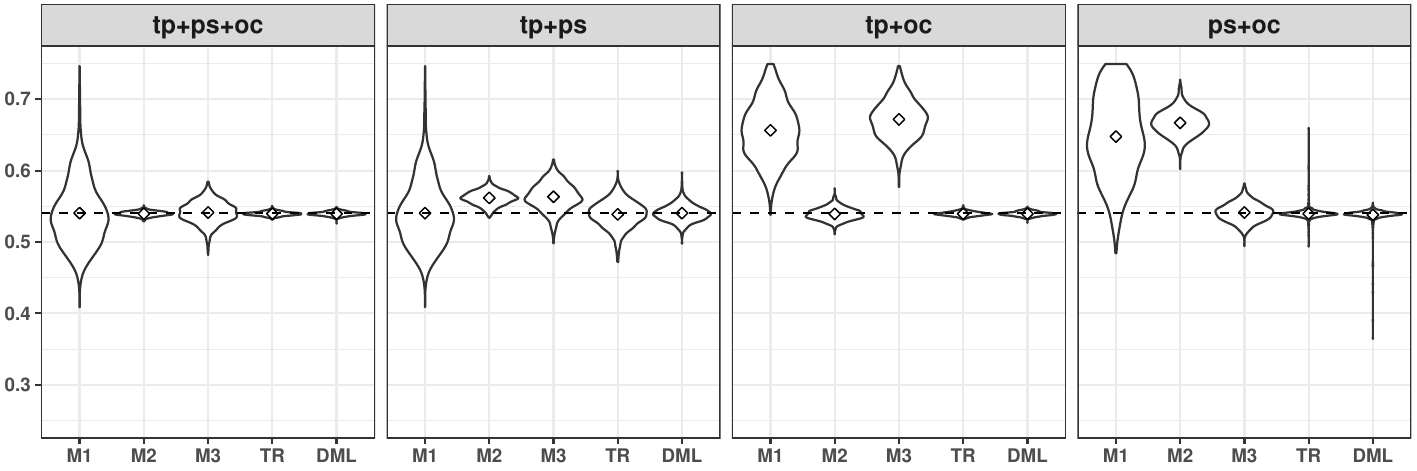}
			\caption{$\tau_h^{11}$}
			\label{subfig:mw-11}
		\end{subfigure}
		\caption{Violin plots of estimators for the principal probabilistic index estimand in scenarios where at most one nuisance function is misspecified. Labels: `tp' for treatment propensity or propensity score; `ps' for principal score; `oc' for pairwise outcome mean; `tp+ps+oc' for the scenario where all three nuisance functions are correctly specified, and `tp+ps' for the scenario where the treatment propensity (propensity score) and the principal score are correctly specified, etc; `M1', `M2', and `M3' for plug-in moment estimators $\widehat{\tau}_{h,\pi+p}^s$, $\widehat{\tau}_{h,\pi+\mu}^s$, and $\widehat{\tau}_{h,p+\mu}^s$, respectively; `TR' for the triply robust estimators given in \eqref{eq:eif-estimator}; and `DML' for the debiased machine learning estimators given in \eqref{eq:eif-estimator-ml}. Horizontal dashed lines represent the true values of the PGCEs.}
		\label{fig:sim-results-mw}
	\end{figure}

	For the plug-in moment estimators and the triply robust estimator, treatment propensity scores and principal scores are estimated using logistic regression, whereas the pairwise outcome mean is estimated using linear regression, as in Example \ref{example-mwp}. For the debiased machine learning estimator, we adopt a five-fold cross-fitting ($K=5$ and $L=15$) and estimate nuisance functions using the \texttt{SuperLearner} with the random forest and generalized linear model libraries. We set the number of observations to $n=2,000$ and consider $1,000$ replications. Violin plots containing the simulation results under each considered estimator are given in Figure \ref{fig:sim-results-mw} and Web Figure S1 in Section S6 of the Online Supplement.
	
	Figure \ref{fig:sim-results-mw} summarizes the results of the five estimators in scenarios where, at most, one of the three nuisance functions is misspecified. Plug-in moment estimators are consistent whenever all relevant nuisance functions are correctly specified. In contrast, the triply robust and debiased machine learning estimators are consistent across all four scenarios, regardless of whether all nuisance functions or subsets of nuisance functions are correctly specified. Results of each considered estimator in scenarios where at least two of the three nuisance functions are misspecified are presented in Web Figure S1 in Section S6 of the Online Supplement. As expected, the plug-in moment estimators are generally biased in those scenarios. In contrast, although biased, the triply robust and debiased machine learning estimators overall exhibit improved estimation accuracy. Finally, compared with the triply robust estimator, the debiased machine learning estimator yields a slight improvement because it is generally less biased.
	
	\subsection{Win ratio with an intermediate variable} \label{sec:sim-wr}
	
	We further consider the case of estimating the win ratio estimands introduced in Examples \ref{ex-win} and \ref{example-wr}, with the final potential outcomes being ordinal. We generate covariates $\bfX_i$, the treatment assignment $Z_i$, and the intermediate variable $D_i$ from the same process given in the previous simulation study. However, for the potential outcome $Y_i(z)\in\{1,2,3\}$, we consider the following three-category ordinal logistic regression model:
	\begin{align} \label{eq:sim-wr}
		\log\frac{\bbP\{Y_i(Z_i)\leq q\}}{\bbP\{Y_i(Z_i)>q\}}=\zeta_{q,0}+2Z_i-D_i+X_{i1}-X_{i2}+1.2X_{i3}-0.8X_{i4},
	\end{align}
	for $q\in\{1,2\}$, where $\zeta_{1,0}=-1$ and $\zeta_{2,0}=1$. Transformed covariates are created following Section \ref{sim-PI}. For the plug-in moment estimators and the triply robust estimator with parametric working models, the pairwise outcome mean is estimated using ordered logit regression, while treatment propensity and principal scores are estimated using logistic regression. For the debiased machine learning estimator, we used multinomial generalized linear models with the elastic net penalty from the \texttt{glmnet} package in \texttt{R} to estimate the pairwise outcome mean function. The treatment propensity and principal scores were estimated using the \texttt{SuperLearner}. The debiased machine learning estimator employs a five-fold cross-fitting procedure with $K=5$ and $L=15$. We set the number of observations $n=2,000$ with $1,000$ replications. In parallel to Section \ref{sim-PI}, violin plots summarizing the simulation results of each considered estimator are given in Web Figures S2 and S3 in Section S6 of the Online Supplement.
	
	From Web Figure S2, we observe that the plug-in moment estimators are consistent whenever all relevant nuisance functions are correctly specified. In contrast, the triply robust and debiased machine learning estimators are consistent in all four scenarios where at most one of the three nuisance functions is misspecified. When at least two of the three nuisance functions are misspecified (Web Figure S3 in Section S6 of the Online Supplement), all considered estimators are biased, with the triply robust and debiased machine learning estimators outperforming the others in terms of estimation accuracy (bias).
	
	\subsection{Additional simulations under non-monotonicity} \label{subsec:sim-sa}
	
	We conduct a third simulation study to demonstrate bias in the uncorrected triply robust and debiased machine learning estimators, and to validate the bias-corrected estimators under non-monotonicity. We modify the data-generating processes in Sections \ref{sim-PI} and \ref{sec:sim-wr} by integrating the sensitivity function $\eta(\bfX)$. The detailed data-generating models and numerical results are presented in Section S7.2 of the Online Supplement. Overall, the results confirm that the bias-corrected estimators give consistent estimates of all four PGCEs across scenarios, suggesting their usefulness when monotonicity cannot be assumed.
	
	\section{Data example} \label{sec:app}
	
	We illustrate the proposed methods with a data example from the U.S. Jobs Corps study \citep{Schochet2001}, in which individualized academic education, vocational training, counseling, and job placement assistance were offered at Jobs Corps centers to economically disadvantaged youth aged 16 to 24. The data were from a randomized controlled trial conducted at 119 Job Corps centers across the U.S., where eligible youth were randomly assigned to either the treatment group, which received an immediate offer to participate in Job Corps, or the control group, which did not participate in Job Corps until three years later. As in many open-label randomized controlled trials, substantial noncompliance with treatment assignment occurred as an intermediate outcome 1 year after randomization: 15.4\% ($857/5,577$) of participants assigned to treatment did not adhere to the assignment, and 50.6\% ($1,854/3,663$) assigned to control switched to the treatment condition. 
	
	The proposed methods were implemented to examine the causal effect of the Job Corps intervention, with weekly earnings one year after the assignment selected as the final outcome. As earning outcomes are often skewed and non-normal, studying alternative estimands that are not based on the mean contrast may provide complementary insights into the treatment effects. In each data analysis below, we adjust for baseline covariates, including demographics (e.g., age, gender, ethnicity, etc.), education level, receipt of public assistance, involvement with the criminal justice system, crimes committed against Job Corps participants, use of legal and illegal substances, health and mortality, family formation and child care, and perform principal stratification analysis under the assumption that mean principal ignorability holds conditional on these covariates.
	
	\subsection{SPI for continuous outcomes} \label{subsec:jobs-spi}
	
	We first implement the proposed methods by considering the job status one year after the assignment as the intermediate outcome, $D$, where if participant $i$ is employed one year after the assignment, then $D_i=1$ and the final outcome $Y_i$, the weekly earning, can be observed and defined without ambiguity; otherwise, $D_i=0$ and $Y_i$ is not well-defined and denoted by $Y_i=*$. In other words, we focus on the initial random assignment (an intention-to-treat analysis) and address earnings outcomes truncation by ``death'' (employment status); see also \citet{Zhang2008} for a similar setup. Under this setting, we estimate the SPI estimand defined in \eqref{eq:SPI} for the always-employed (principal stratum $11$), i.e., participants who will be employed with or without Jobs Corps. For methods based on parametric working models, a logistic model is fitted for the employment process within each assignment group, and a log-normal model is fitted for the final earnings outcome to estimate the pairwise mean outcome function. For the debiased machine learning approach, we use a five-fold cross-fitting ($K=5$ and $L=15$) and estimate nuisance functions using the \texttt{SuperLearner} with the random forest and generalized linear models libraries. 
	
	The estimated results for SPI are presented in Table \ref{tab:SPI}, where $1,000$ bootstrap replicates were generated to estimate the standard errors. We observe that, for the always-employed (72.9\% of all participants, by estimation), the probabilistic index is estimated at $0.539$ using the triply robust and debiased machine learning methods. This indicates that the probability that a randomly chosen always-employed under assigned treatment has a higher-ranked earning outcome than a randomly chosen always-employed under control is $0.539$, with the 95\% confidence interval excluding the null of $0.5$; this shows that the Jobs Corps positively affects one's earnings in the one year after the assignment. Finally, the point estimates obtained under the plug-in moment methods are slightly closer to the null compared to those from the triply robust and debiased machine learning methods, although their 95\% confidence intervals continue to exclude the null. 

	\begin{table*}[htbp]
		\caption{Estimated SPI for the Jobs Corps data. `M1', `M2', and `M3' for plug-in moment estimators $\widehat{\tau}_{h,\pi+p}^s$, $\widehat{\tau}_{h,\pi+\mu}^s$, and $\widehat{\tau}_{h,p+\mu}^s$, respectively; `TR' for the triply robust estimators given in \eqref{eq:eif-estimator}; and `DML' for the debiased machine learning estimators given in \eqref{eq:eif-estimator-ml}. The corresponding 95\% confidence interval (CI) is given in the following parentheses.} \label{tab:SPI}
		\centering
		\begin{tabular}{ccc}
			\hline
			Methods & \multicolumn{2}{c}{SPI estimates} \\
			\hline
			M1 & 0.521 & (0.508, 0.533)\\
			M2 & 0.518 & (0.505, 0.530)\\
			M3 & 0.525 & (0.513, 0.538)\\
			TR & 0.539 & (0.524, 0.555)\\
			DML & 0.539 & (0.527, 0.552)\\
			\hline
		\end{tabular}
	\end{table*}
	
	\subsection{Principal win ratio (PWR) for ordinal outcomes} \label{subsec:jobs-pwr}
	
	We provide additional analysis of the Job Corps data to illustrate the application of the proposed methods to address ordinal outcomes with noncompliance. We use the actual treatment status as an intermediate variable, similar to \citet{Chen2015}. Under the monotonicity assumption, three strata potentially exist, including the compliers (stratum $10$), the never-takers (stratum $00$), and the always-takers (stratum $11$). To address the possibility of unemployment status at the time of earning assessment, we create an ordinal outcome with three ordered categories: 0 (unemployment), 0 to 170 per week (hourly wage 4.25 dollars on a 40-hour-per-week basis), and above 170 dollars per week, following the categorization considered by \citet{Lu2018}. Under this setting, we estimate the principal win ratios within each principal stratum, as defined in \eqref{eq:CWR}. For methods based on parametric working models, the pairwise outcome mean is estimated using ordered logit regression, while treatment propensity and principal scores are estimated using logistic regression. For the debiased machine learning approach, we used multinomial generalized linear models with an elastic net penalty to estimate the pairwise outcome mean. The treatment propensity and principal scores were estimated using the \texttt{SuperLearner}, fitted via a five-fold cross-fitting ($K=5$ and $L=15$) procedure. 
	
	The estimated results are presented in Table \ref{tab:CWR}, where $1,000$ bootstrap replicates were generated to estimate the standard errors. The estimated proportions of always-takers, compliers, and never-takers among all participants are 50.1\%, 34.0\%, and 15.9\%, respectively. The Jobs Corps intervention assignment has a positive effect on participants' earnings one year after the assignment for both compliers and always-takers, and this effect appears to be more pronounced for compliers. For example, using the debiased machine learning estimator, there is a 27\% increase in the probability of being in a higher-earning category among compliers under treatment relative to those under control. The 95\% lower confidence limits obtained from all five methods are strictly above the null value of 1. For the never-takers, however, the effect of Jobs Corps appears more neutral, as the principal win ratio is estimated to be closer to 1, and three out of five methods (M1, triply robust and debiased machine learning) produce 95\% confidence intervals that include the null value of 1. Interestingly, in this reanalysis using a win ratio estimand, we find that treatment effect heterogeneity may exist across the three compliance strata, and the exclusion restriction assumption often invoked for compliance analysis may be questionable. The statistically significant effect among the always-takers indicates a positive psychological effect due to the assignment alone. That is, participants assigned to the Jobs Corp intervention may feel encouraged and naturally be more optimistic about future employment opportunities (hence more likely to move into a higher-earning category) than those who switched from their initial control assignment.

	\begin{table*}[htbp]
		\caption{Estimated win ratios within different compliance strata for the Jobs Corps data. `M1', `M2', and `M3' for plug-in moment estimators $\widehat{\tau}_{h,\pi+p}^s$, $\widehat{\tau}_{h,\pi+\mu}^s$, and $\widehat{\tau}_{h,p+\mu}^s$, respectively; `TR' for the triply robust estimators given in \eqref{eq:eif-estimator}; and `DML' for the debiased machine learning estimators given in \eqref{eq:eif-estimator-ml}. The corresponding 95\% confidence interval is given in the following parentheses.} \label{tab:CWR}
		\centering
		\begin{tabular}{ccccccc}
			\hline
			& \multicolumn{6}{c}{Strata} \\
			\cline{2-7}
			Methods & \multicolumn{2}{c}{$10$} & \multicolumn{2}{c}{$00$} & \multicolumn{2}{c}{$11$} \\
			\hline
			M1 & 1.216 & (1.128, 1.304) & 1.096 & (0.955, 1.238) & 1.117 & (1.016, 1.218)\\
			M2 & 1.206 & (1.122, 1.289) & 1.087 & (1.025, 1.149) & 1.096 & (1.025, 1.168)\\
			M3 & 1.204 & (1.121, 1.286) & 1.089 & (1.025, 1.152) & 1.096 & (1.025, 1.167)\\
			TR & 1.197 & (1.112, 1.282) & 1.026 & (0.920, 1.133) & 1.131 & (1.040, 1.223)\\
			DML & 1.270 & (1.146, 1.393) & 1.041 & (0.940, 1.142) & 1.167 & (1.082, 1.252)\\
			\hline
		\end{tabular}
	\end{table*}

	\subsection{Sensitivity analyses for monotonicity}

	We conducted additional sensitivity analyses to examine the impact of potential violations of the monotonicity assumption, focusing on both Sections \ref{subsec:jobs-spi} and \ref{subsec:jobs-pwr}. Following the approach in Section \ref{subsec:sim-sa}, we specified the sensitivity function as $\eta(\bfX)=\eta_0[1-\{p_1(\bfX)-p_0(\bfX)\}/\min\{p_1(\bfX),1-p_0(\bfX)\}]$, with $\eta_0 \in \{0.1, 0.2, 0.3, 0.4, 0.5\}$. This formulation allows the sensitivity function to naturally vary with covariates while being governed by a one-dimensional strength parameter $\eta_0$. The chosen range of $\eta_0$ represents mild to moderate departures from monotonicity, which is plausible in the context of the Jobs Corps study, where encouragement for employment and subsequent treatment receipt may not deviate dramatically from monotonicity. We applied both the triply robust and the debiased machine learning estimators, with results reported in Web Figure S5 in Section S7.2 of the Online Supplement.
	
	For the SPI analysis (Web Figure S5(a)), increasing $\eta_0$, corresponding to stronger departures from monotonicity, led to slightly larger estimated PGCEs. In other words, as the proportion of harmed strata increases, the estimated SPI shifts further from the null, yielding a stronger treatment effect signal, although statistical significance remains unchanged. In contrast, for the PWR analyses (Web Figures S5(b) - (d)), the estimated PGCEs are largely stable across values of $\eta_0$. Despite the presence of defiers, the estimated win ratios across all three principal strata remain nearly unchanged. The only noticeable pattern is a narrowing of the 95\% confidence interval for the complier win ratio as $\eta_0$ increases. These findings suggest that the PWR estimators exhibit strong robustness to non-monotonicity, in contrast to the slightly more pronounced sensitivity observed for the SPI.

	\section{Discussion} \label{sec:disc}

	We develop principal stratification methods for addressing a general class of PGCE estimands defined in terms of a binary intermediate variable. Under the assumptions of treatment unconfoundedness, monotonicity, and mean principal ignorability based on the contrast function, we proposed a semiparametric approach for estimating the PGCEs by identifying their efficient influence functions and obtaining scaled U-statistic-based estimators. We showed that the proposed U-statistic-based estimators are triply robust and locally efficient, with asymptotic variances that attain the semiparametric efficiency lower bounds when all nuisance functions are correctly specified by parametric or machine learning models with cross-fitting. 
	
	As in all principal stratification analyses that achieve point identification, the proposed U-statistic-based estimators of PGCEs also depend on the mean principal ignorability assumptions, which are unverifiable from the observed data alone. In Section S7 of the Online Supplement, we describe an additional sensitivity-function approach to obtain bias-corrected U-statistic-based estimators of the PGCEs when the mean principal ignorability assumption is violated. The bias-corrected estimators remain triply robust to potential working model misspecifications and provide a means to generate a range of point estimates under the assumption of non-ignorability. 
	
	Finally, our current development assumes that the observed data vectors are independent and identically distributed, providing a natural starting point. Extending these ideas to correlated data structures is an important direction for future research. For clustered data, recent work on cluster randomized trials has emphasized the need to define estimands that explicitly account for the multilevel structure, particularly in the presence of informative cluster size \citep{kahan2023estimands}. A promising avenue is to expand the semiparametric and nonparametric methods of \citet{wang2024model} to accommodate cluster-average and individual-average PGCE estimands when the intermediate variable is measured at the individual level. For longitudinal data, both intermediate and final outcomes may be repeatedly measured over time, which requires reformulating the identification assumptions to incorporate sequential ignorability of treatment and multilevel monotonicity \citep{frangakis2004methodology}. It is also worth noting that, while our main exposition has focused on a single final outcome, our framework naturally extends to multiple hierarchically ordered outcomes, a setting that has gained considerable traction in clinical research over the past decade. In Section S8 of the Online Supplement, we extend the notation of \citet{Bebu2015} to the potential outcomes framework and show how our results apply to this setting by defining pairwise comparison estimands with hierarchical outcome components while maintaining the same binary intermediate variable. Further methodological developments along these directions will continue to broaden the applicability of our framework.
	
	\section*{Acknowledgement}
	
	Research in this article was partially supported by the United States National Institutes of Health (NIH), National Heart, Lung, and Blood Institute (NHLBI, grant number 1R01HL178513). All statements in this report, including its findings and conclusions, are solely those of the authors and do not necessarily represent the views of the NIH.
	
	\singlespacing
	\bibliographystyle{jasa3}
	\bibliography{CUSPS}

\end{document}